\documentclass[prb,preprint]{revtex4-1} 

\usepackage{natbib}
\usepackage{amsmath}  
\usepackage{amsfonts} 
\usepackage{graphicx} 
\usepackage{enumitem}
\usepackage{hyperref}
\usepackage{todonotes}

\begin{document}


\title{Using Virtual Reality in Electrostatics Instruction: \\ The Impact of Training}

\author{C. D. Porter}
\email{porter.284@osu.edu} 
\altaffiliation[permanent address: ]{191 W. 19th Ave, 
  Columbus, OH, USA} 
\author{J. R. H. Smith}
\author{E. M. Stagar}
\author{A. Simmons}
\author{M. Nieberding}
\author{C. M. Orban}
\affiliation{Department of Physics, The Ohio State University, Columbus, OH 43210}

\author{J. Brown}
\email{brown.4972@osu.edu}
\affiliation{Department of Engineering Education, The Ohio State University, Columbus, OH 43210}

\author{A. Ayers}
\email{ayers.224@osu.edu}
\affiliation{Advanced Computing Center for the Arts and Design, The Ohio State University, Columbus, OH 43210}


\date{\today}

\begin{abstract}
Recent years have seen a resurgence of interest in using Virtual Reality (VR) technology to benefit instruction, especially in physics and related subjects. As VR devices improve and become more widely available, there remains a number of unanswered questions regarding the impact of VR on student learning and how best to use this technology in the classroom. On the topic of electrostatics, for example, a large, controlled, randomized study performed by Smith et al. 2017\cite{smith17}, found that VR-based instruction had an overall negligible impact on student learning compared to videos or images. However, they did find a strong trend for students who reported frequent video game play to learn better from VR than other media. One possible interpretation of this result is that extended videogame play provides a kind of ``training" that enables a student to learn more comfortably in the virtual environment. In the present work we consider if a VR training activity that is unrelated to electrostatics can help prepare students to learn electrostatics from subsequent VR instruction. We find that preliminary VR training leads to a small but statistically significant improvement in student performance on our electrostatics assessment. We also find that student reported game play is still correlated with higher scores on this metric.
\end{abstract}

\maketitle 

\section{Introduction} 

Many topics in physics are inherently three-dimensional (3D), but are usually taught using two-dimensional media such as whiteboards and computer screens. Stereoscopic virtual reality (VR) allows students to view 3D scenes with depth perception, which should be advantageous for teaching certain content in physics and other STEM disciplines. Efforts to develop stereoscopic VR visualizations for physics began in the mid-1990s \cite{ScienceSpace,ScienceSpace2,ScienceSpace3} and continue to the present day (e.g. \cite{Pirker_etal2017,Hu_etal2018,Liu_etal2019,Reed2019,Greenwald_etal2019,PERCVR2019} and references therein) as the technology improves.

The growth in VR\footnote{In this article we do not say much about Augmented Reality (AR). While AR technology is maturing rapidly and holds great potential for classroom impact, it remains true that AR technology that provides stereoscopic depth perception is still very expensive and not within access of most instructors.} content for physics should be followed by detailed studies of the impact of these visualization methods on student learning. The studies that have been performed in physics and related STEM fields report varying degrees of success \cite{osuVR,john5,will4,zhou11,mor4,winn92,kauf00,kauf03,trin02,madd18,madd19}, including many cases in which stereoscopic visualization techniques did not prove to be pedagogically more valuable than more conventional visualization methods.
In this paper we will present data from a new study in a large introductory electromagnetism class at Ohio State University that will address these questions. A particularly affordable way to provide students with a reasonably high-quality VR experience that we emphasize in this paper is so-called Google Cardboard \cite{cardboard} in which a typical smartphone is placed in a cardboard or plastic headset which may only cost a few dollars. This reduced cost is important because it means each student can potentially have their own VR headset, so that VR can become a regular part of instruction. The reduced cost allowed us to perform a large study using a set of six affordably priced smartphones.

Prior studies investigating the effectiveness of VR in physics and astronomy \cite{osuVR,john5,will4,zhou11,mor4,winn92,kauf00,kauf03,trin02,madd18,madd19,Greenwald_etal2019,Blanco_etal2019} have yielded mixed results.
Although students given VR interventions often report being more engaged with the material, and physical immersion in VR has been shown to increase spatial awareness in search tasks\cite{pausch97}, the advantage of VR over other media in achieving gains in specific learning outcomes is still unclear. Unfortunately, because of the prohibitive cost of conventional VR headsets, many of these prior studies have limited sample sizes and in some cases VR treatment was not compared to a control group. 

There have been a few large studies with careful controls. Madden et al. \cite{madd18} considers a VR intervention for an astronomy course on the topic of the moon phases. That study had 172 participants across three treatment groups (VR, computer, and ``hands-on"/control), and found no statistically significant difference in learning gains between treatment groups. A fuller description of their study appears in \cite{madd19}.

Other large studies by Smith et al.\cite{smith17} and Porter et al.\cite{PERC19}, which include several authors of this paper, did not find statistically significant differences in pre-post test gains for VR compared to other media on topics of electrostatics and magnetostatics. The studies involved, respectively, 301 and 289 participants from college-level introductory physics classes. Also of note is a study by Greenwald et al.\cite{Greenwald_etal2019b} where 20 college students completed activities relating to electrostatics and answered questions. Students who completed the activities in a VR headset and interacting with the virtual environment did not outperform students who completed essentially the same activities by drawing on a tablet (i.e. a 2D medium).

If one looks outside the physics content areas just described, there do exist large studies in which statistically significant effects of stereoscopic VR compared to other media have been detected. A notable example from college-level mathematics is Porter and Snapp\cite{porter19}. A recently published meta-study by Merchant et al.\cite{Merchant2019} considered dozens of K12 VR studies and found (among other conclusions) that VR content overall tends to be effective in producing learning gains. However, the goal of Merchant et al.\cite{Merchant2019} was not to weigh the usefulness of stereoscopic VR versus more traditional media, and the meta-study considered non-stereoscopic virtual worlds accessed through conventional desktop and laptop computers to be VR. So while the paper is very interesting and thorough, its relevance is in many ways oblique to the work that we will describe here.

In Smith et al. \cite{smith17}, although VR did not prove to be more effective for students in general, it was found that students who reported frequent video game play (a.k.a. ``gamers") and were given the VR treatment had much higher gains than any other group (non-gamers, or gamers who received electrostatics instruction from video renderings or images). Porter et al. found a similar trend for ``gamer" students to significantly outperform non-gamer peers on magnetostatics assessments after viewing magnetostatics content, although, interestingly, the VR treatment did not help the gamer students more than other media as Smith et al. found. Both Smith et al. and  Porter et al. interpret these results to imply that gamers may have a ``familiarity" with visuospatial rotations to the point where they are less likely to be cognitively overwhelmed by and better able to learn from inherently 3D content. Smith et al.\cite{smith17} and  Porter et al.\cite{porter19} both conclude with questions asking if ``repeated exposure" to VR or ``training" of students with VR can help non-gamers learn as effectively as gamers do from VR content.

Building off of this line of inquiry, in the present study we consider if VR training activities on topics unrelated to electrostatics can improve students' ability to learn effectively from VR instruction on electrostatics. These activities are described in the next section. The inclusion of virtual training prior to engaging in this experience has been utilized across other virtual environments, such as the “landing room” developed by Johnston et al.\cite{johnston18} This space forms a familiar environment for users to learn necessary interactions and adjust to a new visual format before diving into an abstract view of a cell membrane. While other examples of this training exist in other studies, few have demonstrated its impact on overall performance.

Due to a lack of independently-validated assessments for electrostatics with a high fraction of 3D questions, we developed a suite of questions as a preliminary survey of 3D electrostatics (see the Methods section). This is only briefly summarized there because of page constraints. The reliability of this survey is discussed below, along with student performance.


\section{Methods}

The subjects of this work were students in the second semester of an introductory calculus-based physics course at a large Midwestern university, offered in autumn. This course was being offered ``off-sequence" meaning that students who begin physics in their first semester would have taken this course in spring. Students were offered the equivalent of one homework assignment's course credit for coming to our lab and participating in either the research study, or an alternative assignment of roughly equivalent length. Of 281 initial respondents, 279 agreed to participate in research.
    
As students entered the testing area, they were randomly assigned to one of two treatment types: VR with preliminary training, and VR with no initial training. The assessments were identical for all students, regardless of treatment type, except for a few questions posed during the preliminary training, which were unrelated to electrostatics. The students' average overall performance in physics was fairly constant between treatment types, as determined by post-hoc analysis of students' final scores as a percentage of points in their physics course 

(Training: $(84.5 \pm 0.9)  \%$, No Training: $(84.2 \pm 0.9)  \%$, $p>0.8$). There was almost no variation in the percentage of students reporting their sex as female in the two treatment types (Training: 20\%, No Training: 20\%). Although gender identity would be a better descriptor of participants, gender identity is not available.
    
\subsection{Treatments}
    
VR visualizations were created as Android smartphone applications. The apps were written using Unity, a cross-platform game engine developed by Unity Technologies \cite{unity19}, and the Google VR SDK for Unity.  Smartphones were placed in plastic goggles which have lenses to focus the near point of the eye, and a divider to split the field of view. The smartphone displays an app in a split-screen mode so that 3D phenomena are shown on the right half of the phone to the right eye from an angle slightly to the right, and the equivalent is shown to the left eye from an angle slightly to the left. This creates a stereoscopic 3D virtual reality experience giving the impression of depth perception.
    
\begin{figure}[ht]
\centering
\includegraphics[width=0.75\columnwidth]{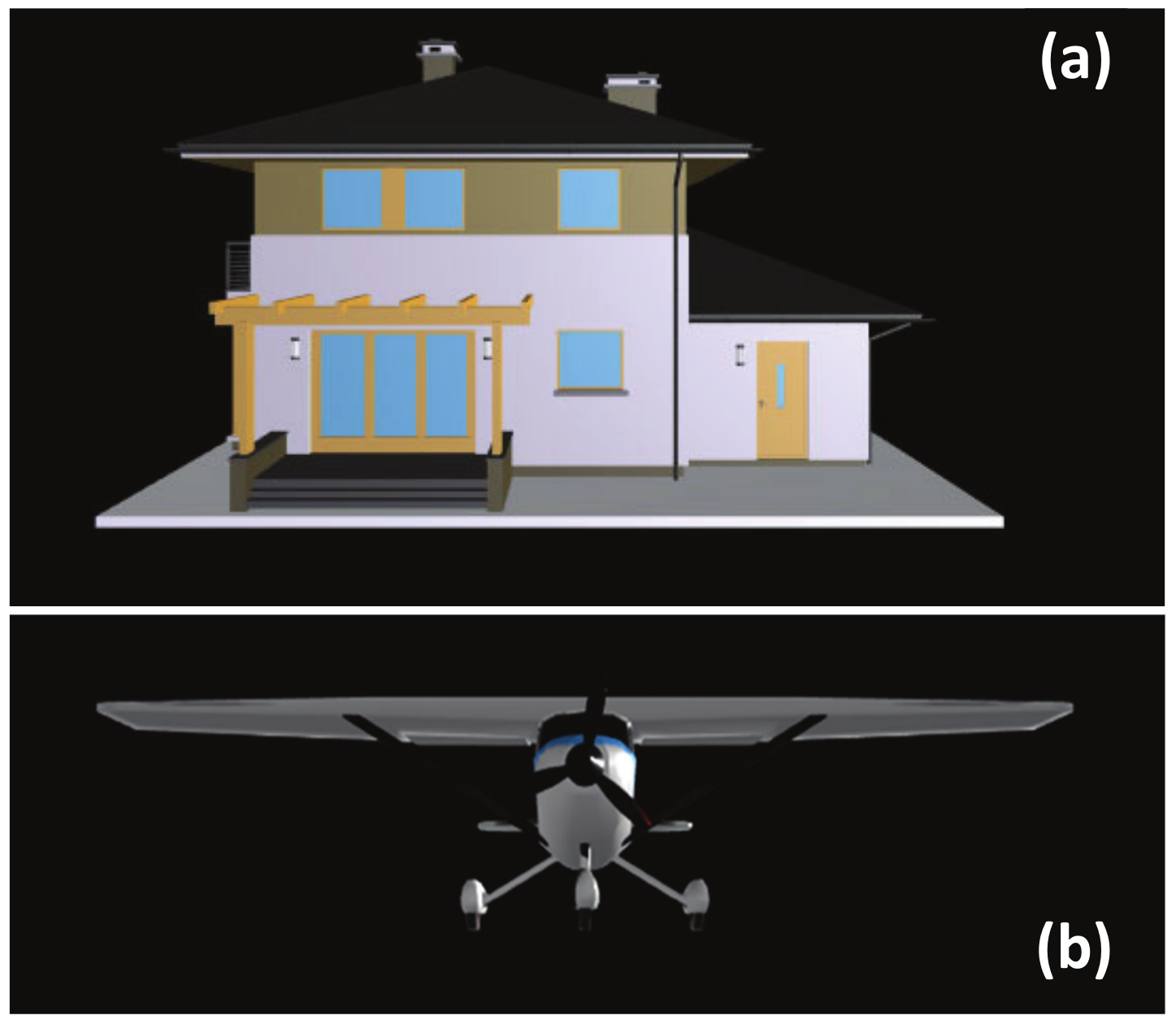}
\caption{Screen shots from two preliminary training scenes. Students were asked to rotate the house in (a) and count the panes of glass. They were then asked to view the rotation of the propeller on the plane in (b) and determine the direction of the angular momentum using the right-hand rule. }
    \label{fig:TrainScenes}
\end{figure}

Preliminary training: Students in the preliminary training group viewed scenes that were unrelated to electrostatics. In the first scene, students were shown a 3D model of a house, and were asked to rotate the house, view it from all angles, and count the number of windows. In the second scene, students were shown a 3D model of a single-propeller airplane, and were asked three questions related to angular momentum such as the initial direction of $\vec{L}$, and the change in $\vec{L}$ if certain maneuvers are performed. Screenshots from these training scenes are shown in Fig. \ref{fig:TrainScenes}.
    
    
\begin{figure}[ht]
    \centering
 \includegraphics[width=1.0\columnwidth]{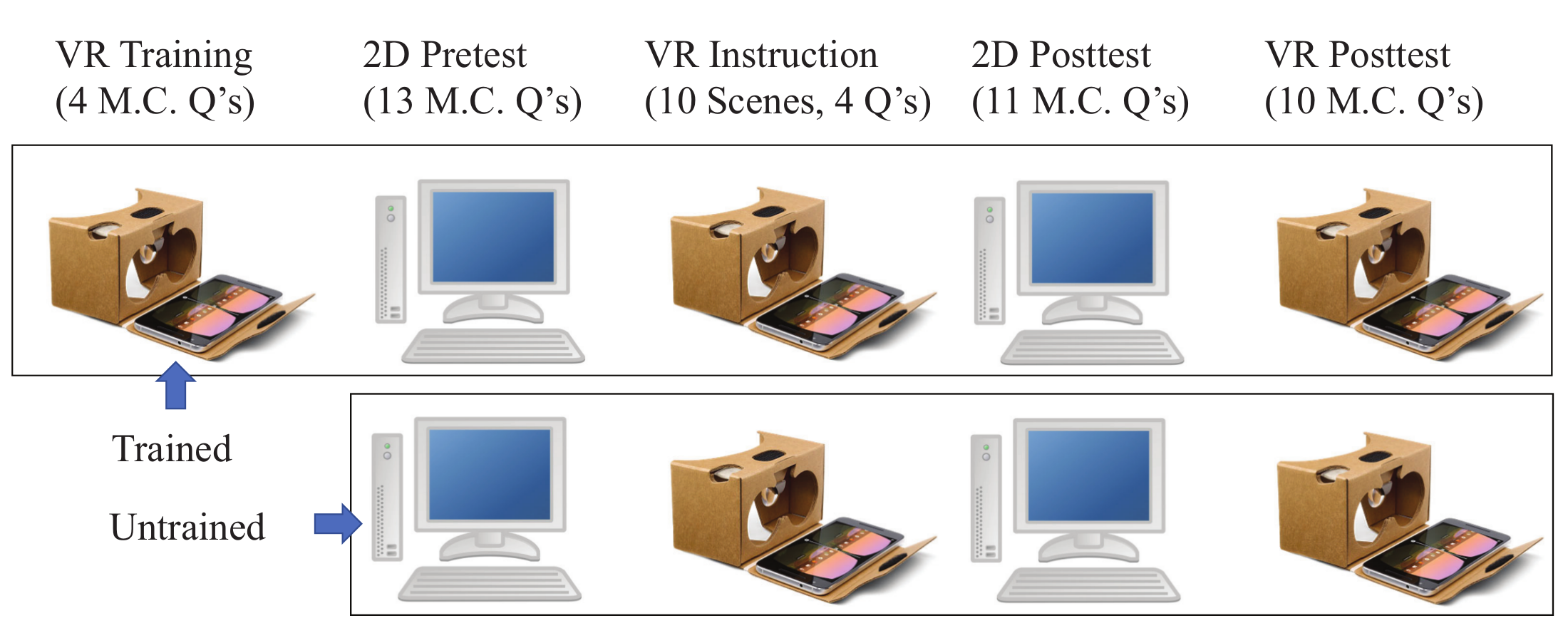}
    \caption{The sequence of treatment and test questions applied to the two groups: those who received preliminary VR training, and those who did not.}
    \label{fig:protocol}
\end{figure}

Only the preliminary training group was given these initial scenes. Students took an average of 4 minutes on all training scenes combined. All students took a pretest on a 2D computer screen with 13 multiple choice questions on electrostatics. Students then moved on to the electrostatics VR instruction which consisted of several visualizations of electric fields, and a few in-VR questions to promote engagement. This instruction is described in greater detail below. Students were then given the posttest (on a computer) consisting of 11 multiple choice questions, followed by 10 in-VR posttest questions. A schematic of the testing protocol is shown in Fig. \ref{fig:protocol}.
    
VR Instruction: The VR scenes on electrostatics shown to both treatment groups involved visualization of electric fields due to charge distributions including an electric dipole, a long, charged rod, and a large, charged plate. The electric field was represented as an array of vectors as opposed to using the density of continuous field lines. The application was then built as an Android application package (APK file), and installed on two Nexus 5X smartphones. The app splits the phone's screen into two halves, one for each eye. Each phone is then placed in a cardboard or plastic viewer. The students can then view the electric systems in stereoscopic 3D. The app utilizes the smartphone sensors so that when the students turn their heads, the system being displayed on the screen rotates, allowing students to see it from any orientation. Students were shown 7 instructional scenes and were told to “look around” and study the magnetic field vectors from many angles before moving on. Students were also asked a series of 3 questions within the VR simulation to ensure that students were engaging with the content. Students controlled the rate at which the visualizations progressed.
    
\subsection{Assessment}

Discussions with experienced instructors were used to determine which aspects of electric fields are commonly prioritized in their learning goals for this course.
The study team then designed a set of problems on this content that are highly three-dimensional in nature, and are therefore most likely to be aided by stereoscopic 3D treatments. These problems fall into two broad categories: (1) determining the direction of $\vec{E}$ at locations that are not simply co-planar with a distribution of charge, and (2) understanding features of the vector field as a whole, such as divergence.
    
\begin{figure}[ht]
    \centering
 \includegraphics[width=1.0\columnwidth]{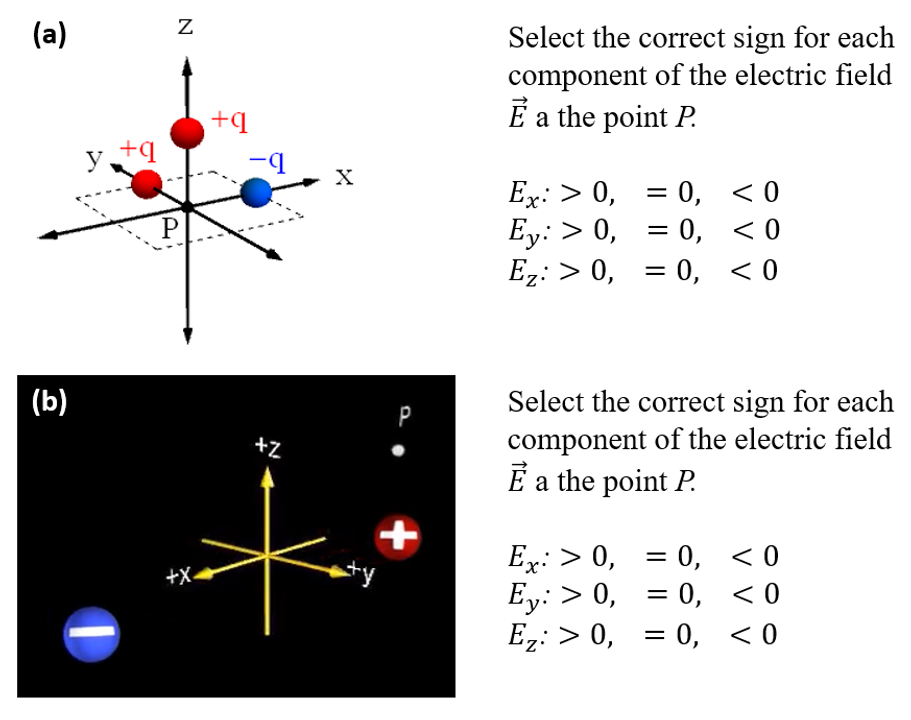}
    \caption{Two example items from the assessment. Item (a) was posed using a 2D computer screen, whereas Item (b) was posed in VR, such that students could rotate the charge arrangement and clearly see that the point P was directly above the positive charge.}
    \label{fig:sample}
\end{figure}

The study team wrote 13 pretest questions and 11 posttest questions to address the above topics. A complete description of the assessment is beyond the scope of this paper. But because these questions have not been independently validated by other groups, we provide in the present work additional statistics on the reliability of this assessment, and some example assessment items in Fig. \ref{fig:sample}.
    
To avoid confounding factors related to medium, reliability analysis was performed only the questions posed using a 2D computer screen. Treating these questions as a preliminary scale for three-dimensional understanding of electric fields, a reliability analysis in SPSS reveals a Cronbach's alpha of 0.91 (or 0.83 for the pretest only, 0.82 for the posttest only). A factor analysis in SPSS revealed a single factor with an eigenvalue greater than 1 (4.8), and this factor explained 43\% of the variance. From these data we conclude that although the assessment still needs to be independently validated and could be improved in many ways, it does appear to be internally consistent and statistically well-behaved.

We note that the questions used on this assessment are not identical to questions used in Smith et al., although there is some overlap. Some questions from Smith et al. were altered to allow for partial credit if students get some Cartesian components of the electric field direction correct, but not all. For example, students answering the item shown in Fig. \ref{fig:sample} would have received 1/3 of a point for each component answered correctly. Questions were also added that included arrangements of three or four charged particles, such that the particles and point at which the electric field direction is to be determined did not lie in any 2D plane (as in Fig. \ref{fig:sample}(a)). These differences, coupled with the fact that students from Smith et al. were from an ``on-sequence" course, mean that these studies cannot be directly compared, quantitatively.

\subsection{Gaming}

One additional difference between the present work and Smith et al. is that in Smith et al. students were asked about gaming frequency, but were not asked about the type of game they primarily play. In this work, students were asked how frequently they currently play videogames, and were then asked whether the games are primarily 2D, 3D, or both. Common examples of each were given (such as Candy Crush for 2D, and Minecraft for 3D). The intent was to classify students based on their responses as either ``3D gamers" or ``Not 3D gamers", since the expectation was that 3D gaming would be key. However, a post-hoc analysis showed no significant differences in our results when students are classified solely on frequency of gaming, with no consideration of 3-dimensionality. This brings into question whether familiarity with 3D visuospatial rotation in an electronic context is truly an explanation of the interaction effect between gaming and gains in Smith et al. and Porter et al. This is discussed further in the Discussion and Conclusions section. This independence of the 3D nature of the games played leads us to classify the groups simply as ``gamers" and ``non-gamers", with ``gamers" being those students who reported playing once per week or more. The present work contains no further comparison between 3D gamers and other types of gamers.

\section{Results and Discussion}

\begin{figure}[ht]
    \centering
 \includegraphics[width=1.0\columnwidth]{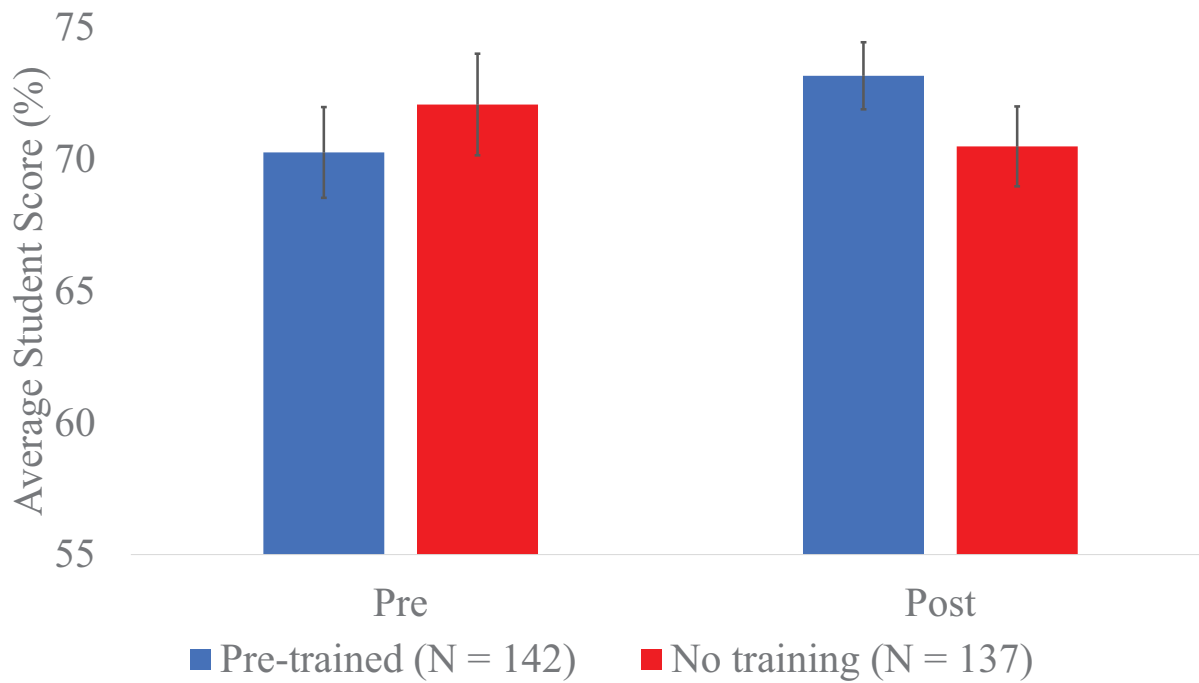}
 \vspace{-3cm}
    \caption{Average scores on pretest and postests for the group that received preliminary training, and the group that did not.}
    \label{fig:ScoreTrain}
\end{figure}

We find that the pre-trained group did have higher gains than the untrained group (see Fig. \ref{fig:ScoreTrain}). The difference in gains is statistically significant with $p=0.014$, and an effect size of $d=0.24$. This result is clearly driven by an increase in scores by the trained group's score and a pre-to-post \emph{decrease} in the untrained group's score. This is consistent with trained students either learning from the intervention, or improving slightly due to the retesting effect. But it is also consistent with the untrained students being overwhelmed by or confused by the intervention, causing lower posttest results than pretest results. Here, pretest questions were all posed on a 2D computer screen. But posttest questions posed in different formats have been grouped together. Some posttest questions were asked entirely on a 2D computer screen and others were posed in VR with answers recorded on a separate computer. These categories of question are compared further below. Questions posed in VR {\it during} the electrostatics instruction have not been counted as either pre or post; they are discussed further below. In all cases, answers were recorded using a 2D computer screen, even if they were posed in VR. These various formats are discussed further below.

\begin{figure}[ht]
    \centering
 \includegraphics[width=1.0\columnwidth]{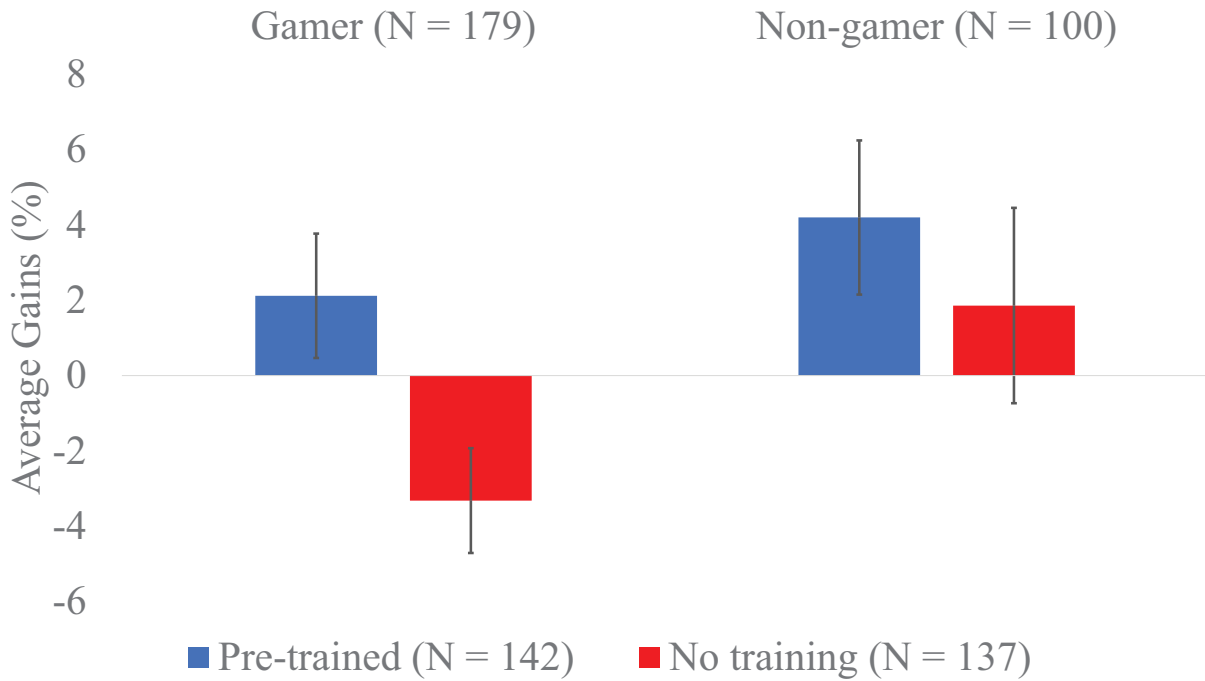}
 \vspace{-3cm}
    \caption{Average post-pre gains for the group given preliminary training, and the group that did not, separated by those reporting high gaming experience (gamer) and low gaming experience (non-gamer).}
    \label{fig:ScoreTrainGame}
\end{figure}

Because this work was initially motivated by the correlation between gains from VR treatments and gaming experience, it is worth breaking down these scores by gaming experience. Fig. \ref{fig:ScoreTrainGame} shows the gains from the two treatment groups broken down by prior gaming experience. 

That trained gamers show positive gains and that trained non-gamers show yet higher gains fits the hypothesis that training can compensate for a lack of familiarity with virtual environments and visuospatial rotations.
However, we were surprised to see that gamers who did not receive preliminary training performed \emph{worse} than any other group, having negative gains. Untrained non-gamers, for example, had small but positive gains. In light of these inconsistent results, it is important to note that there are no statistically significant differences between the performance of gamers and non-gamers in their physics course overall (gamers: 84.0\% $\pm$ 0.8\%, non-gamers: 85.2\% $\pm$ 1.0\%, $p=0.35$). The inconsistent interaction effect between training and gaming casts some doubt over the the simple hypothesis that prior video game play provides important advantages to students for which training can at least partially compensate. It is unclear whether the hypothesis may yet be true, since this inconsistent interaction effect of gaming and training is not statistically significant ($p=0.41$, repeated measures analysis of variance, with treatment and gaming score as between-subjects factors).

\begin{figure}[ht]
    \centering
 \includegraphics[width=1.0\columnwidth]{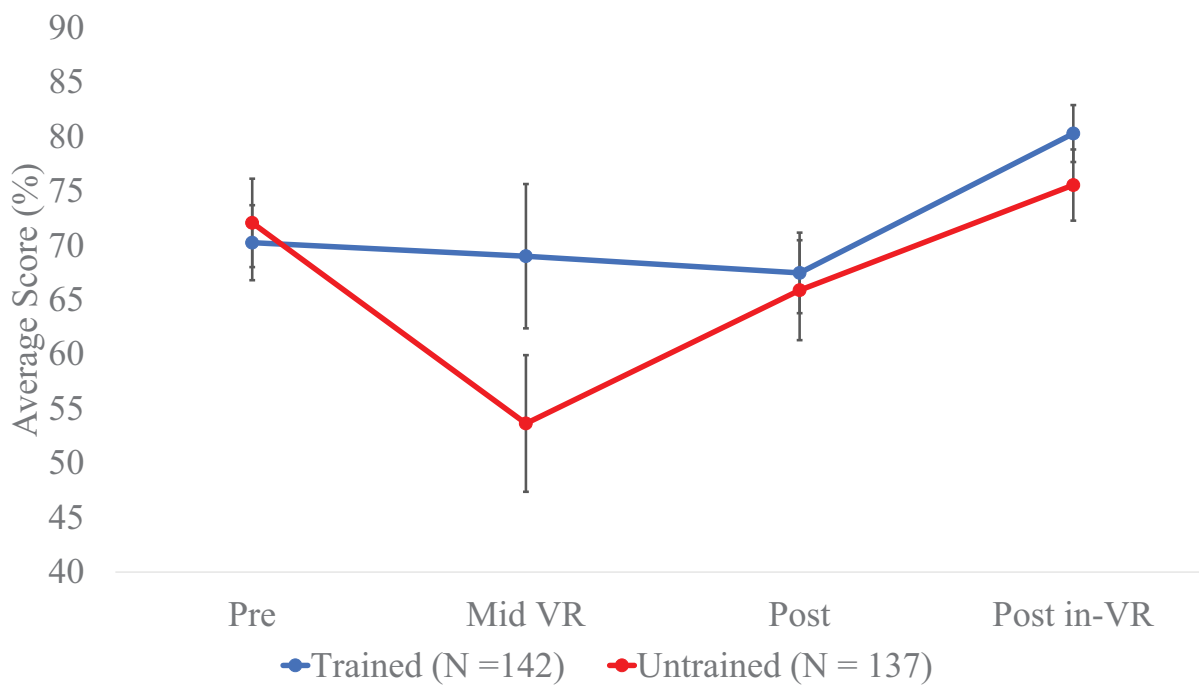}
 \vspace{-3cm}
    \caption{Average scores at different time points on questions posed in different media, separated by treatment type.}
    \label{fig:ScoreTimeTrain}
\end{figure}

To obtain some insight into this paradoxical result, we considered that questions were asked in two formats: some questions were posed in VR while others were posed on a conventional computer monitor (i.e. ``in 2D").
These questions can be categorized into four groups: 1) pretest questions asked entirely on a computer monitor, 2) questions posed in VR during electrostatics instruction, 3) posttest questions asked on a computer monitor, and 4) posttest questions posed in VR. Fig. \ref{fig:ScoreTimeTrain} shows student scores on each of these question sets, arranged in chronological order, and split into trained and untrained groups.

Pretest scores for the two groups (which were asked in 2D) are consistent within standard error. Likewise, the non-VR posttest scores are consistent between the two groups and the overall result is that the non-VR posttest scores are slightly lower.
It is unlikely that the posttest questions are significantly more difficult than the pretest questions, because the two question sets are identical up to occasional swapping of positive and negative charges, or rotation from one high-symmetry point to an analogous high-symmetry point. This being the case, the similarity between the ``pre" and non-VR ``post" test does seem to imply that the VR intervention had a negligible impact on student learning, regardless of whether students were in the trained or untrained group, and contrary to our expectations.

Perhaps the most interesting feature of Fig.~\ref{fig:ScoreTimeTrain} is the large and statistically significant ($d=0.22$, $p=0.02$) difference between the two treatment groups' scores on the questions posed in VR at the midpoint, during electrostatics instruction in VR. The group that received preliminary training in VR performed approximately as well on the questions posed in VR as they did in the pretest (which was in 2D), whereas those who received no initial acclimation to VR had scores about 16\% lower at this mid-point. So in this sense there was a net benefit for students who received the initial acclimation with VR, but the benefit was limited to questions posed in VR.

The additional 7-10\% uptick in scores at the final set of posttest questions posed in VR shown in Fig.~\ref{fig:ScoreTimeTrain} must be viewed with some skepticism, as we cannot verify that the VR-based questions were of comparable difficulty to the other non-VR question sets. The change in medium, for example, is necessarily accompanied by slightly different posing of questions and style. Conceivably, this uptick could be indicative of continued improvement in student understanding of the material and engagement with VR, but such a claim would require significant additional experimental evidence.

\begin{figure}[ht]
    \centering
 \includegraphics[width=1.0\columnwidth]{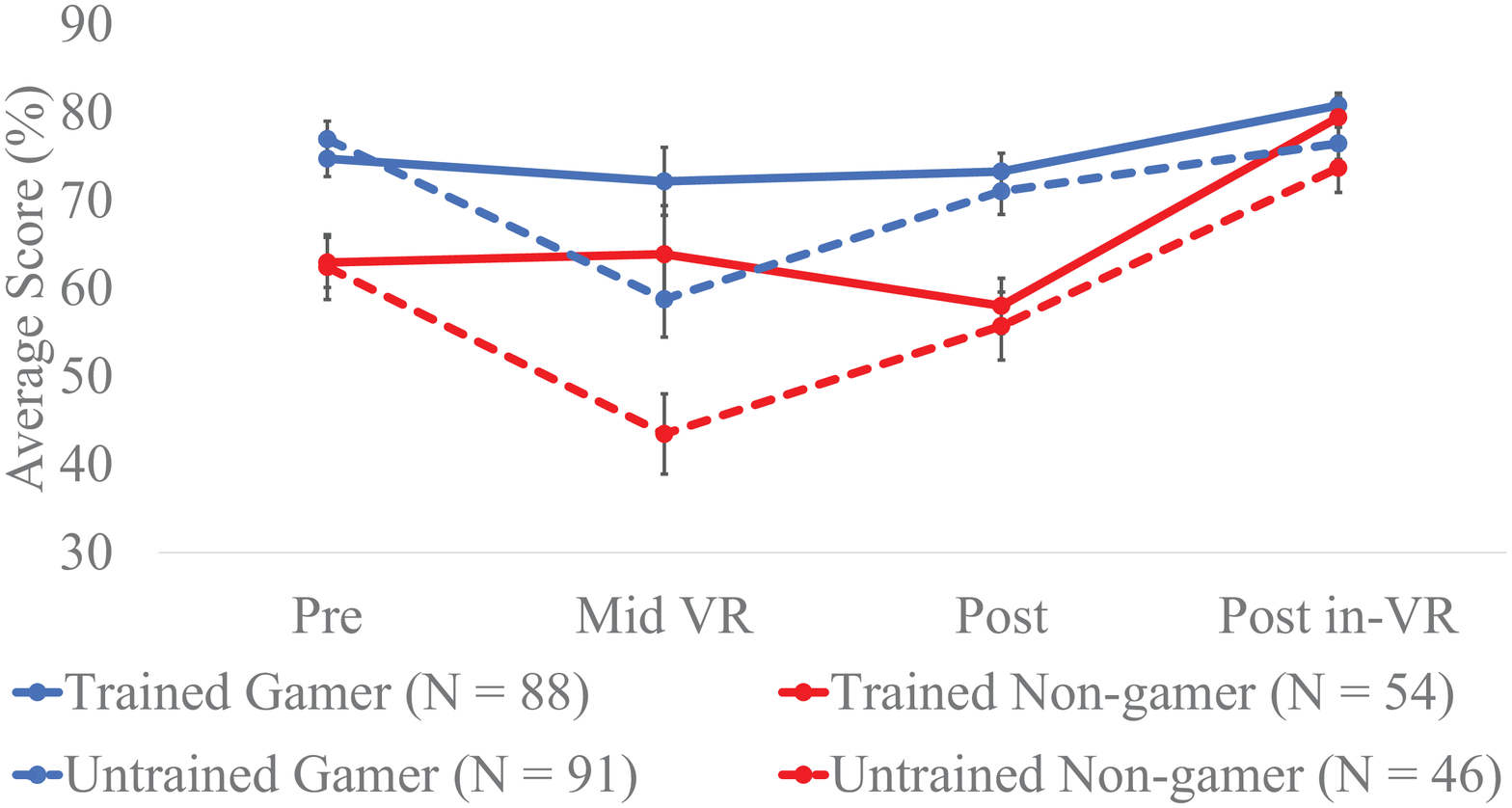}
 \vspace{-3cm}
    \caption{Average scores at different time points on questions posed in different media, separated by treatment type and gaming experience.}
    \label{fig:ScoreTimeTrainGame}
\end{figure}

Given that this inquiry was prompted by an apparent connection of the subject matter with student gaming history, we further break down these four question sets by student gaming history. This is shown in Fig. \ref{fig:ScoreTimeTrainGame}.

The dashed lines show the scores by the untrained group, and the solid lines show those of the trained group. Here we see that the drop for ``Mid VR" questions is seen in both untrained gamers and untrained non-gamers. We also see that, overall, gamers score higher on the pretest than non-gamers. 
Interestingly, by the second set of VR questions, which is the last set of questions that students complete, there is no statistically significant difference between these groups either by gaming or by training. 

\begin{figure}[ht]
\centering
\includegraphics[width=1.0\columnwidth]{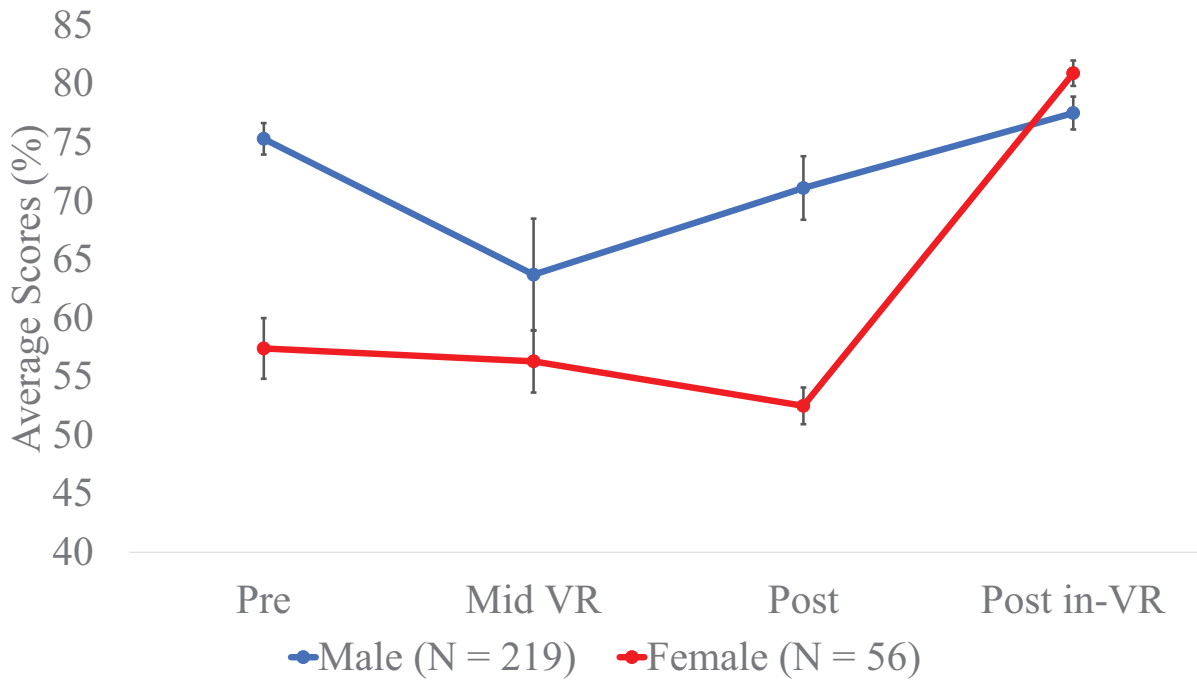}
\vspace{-3cm}
\caption{Average scores at different time points on questions posed in different media, separated by sex. Three students included in other figures are excluded here due to identifiability concerns and/or response completeness.}
\label{fig:ScoreTimeSex}
\end{figure}

It bears mentioning that the non-gamer group has proportionally more women than the gamer group.
As is shown in Fig. \ref{fig:ScoreTimeSex}, this means that equalization between self-reported gamers and non-gamers by the last question set also corresponds to equalization between males and females.

\subsection{Student feedback}

\begin{figure}[ht]
    \centering
 \includegraphics[width=0.8\columnwidth]{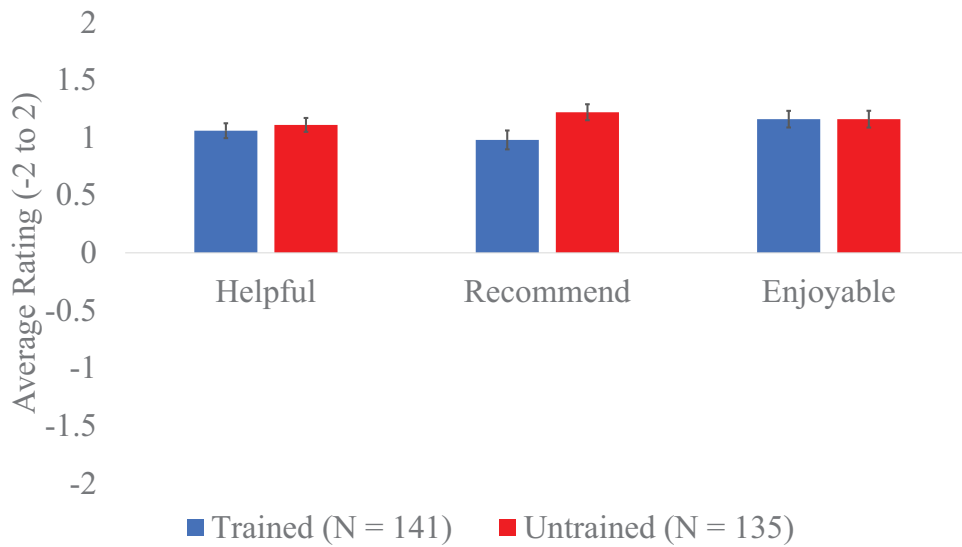}
 \vspace{-2cm}
    \caption{Average feedback on VR intervention separated by treatment group. There were no significant differences between treatment groups in perceived helpfulness of the intervention, its enjoyability, nor in how likely students would be to recommend it to a friend. Three students included in other figures are excluded here due to response completeness.}
    \label{fig:FeedbackTreat}
\end{figure}

Upon completion of all questions related to electrostatics, but prior to being asked demographics questions, students were asked to rate the VR intervention in three ways. They were asked 1) ``How helpful was this intervention?", 2) ``How enjoyable was this intervention?", and 3) ``How likely are you to recommend this intervention to a friend?". Students were asked to respond using a 5-point scale ranging from -2 (``Highly unhelpful", ``Highly unenjoyable", ``Highly unlikely", respectively) to +2 (``Highly enjoyable", etc.) with 0 being neutral. In all cases, average scores were positive, very close to +1 (``Helpful", ``Enjoyable", ``Likely"). Figure \ref{fig:FeedbackTreat} shows that both groups (trained and untrained) found the VR instruction equally helpful and equally enjoyable. Although there was a slight difference in how likely the two groups were to recommend the VR to a friend, the difference is not significant after a post-hoc (Bonferroni) correction.

We find analogous results when splitting student responses according to sex and according to gaming history (not shown). There are no differences in means that are significant after a post-hoc correction, and there are also no large differences in the distributions of responses. These questions were also asked of students in the study by Smith et al. \cite{smith17} which explored student learning comparing VR to other treatment types including videos and still images. Students significantly preferred VR over other treatment types.


\section{Discussion and Conclusion}

Smith et al. \cite{smith17} found that students who reported frequent video game play seem better able to learn from VR-based instruction on electrostatics than ``non-gamers". As discussed in that study, these ``gamer" students who received VR instruction significantly outperformed students who received equivalent instruction from videos or still images. 
In designing the present study our hypothesis was that preliminary training in VR on a topic unrelated to electrostatics would help even non-gamer students perform as well on electrostatics problems as gamer students. 

Figure \ref{fig:ScoreTrain} shows that there is indeed an interaction effect between preliminary training and subsequent gains. It should be noted, however, that the gains averaged over all groups (trained and untrained) are not statistically different from zero.

If comfort with visuospatial rotations in an electronic context (either through training or gaming history) were all that were required to learn effectively from VR, then one would expect there to be a significant dependence on the type of game typically played (2D or 3D). This was not the case. It is possible that frequent gaming of any kind increases the likelihood of some minimum exposure to 3D gaming, even if more time is spent on 2D gaming than on 3D gaming. Greater clarity could be achieved through a simple modification of the gaming experience question, such as asking students to move a slider to indicate the fraction of gaming that is 2D and the fraction that is 3D. Another possibility is that exposure to gaming of any kind increases student self-efficacy in the space of electronic visualization. Self-efficacy, pioneered by Bandura \cite{Bandura77}, is often strongly correlated with perseverance in a discipline or activity. Although this work shows that training can improve gains from VR instruction, the mechanism for this is still uncertain.

Note that in Fig.~\ref{fig:ScoreTrainGame}, students who did report frequent video game play but who did not receive the training improved the least, with an overall worse score on the posttest than on the pretest. This would appear inconsistent with any of the explanations above related to self-efficacy, or accumulated comfort with electronic visuospatial rotations, but the story is clearer in Fig.~\ref{fig:ScoreTimeTrainGame}. There, one can see that the initial average score of untrained gamers was as high (within error bars) as any score achieved by any other group at any time point, such that the small negative gains may be attributable to ceiling effects. Other results on Fig.~\ref{fig:ScoreTrainGame} were unsurprising and generally the training helped students to perform overall better on posttests. 


Figure \ref{fig:ScoreTimeTrain} shows that the primary effect of VR training was to increase the trained group's scores on in-VR questions during instruction. The training, however, did not improve scores on the 2D, computer-based posttest questions. Also, trained and untrained groups achieve nearly the same score on the posttest in-VR questions. One possible explanation of this is that when electrostatic questions are posed \emph{in VR} to students who have never used VR, the experience is overwhelming and they do poorly. But they do better later on when similar questions are asked in VR the second time, such that the midpoint VR experience serves like a preliminary training for the initially untrained students. In other words, all students perform better in VR after one or more exposures to questions in VR. 
The VR instruction appears to have had no effect on the computer-based post test. This indicates either that the in-VR training is ineffective, or that learning in the VR context is not transferring to the 2D context, or both. 

The breakdown of the data in Fig. \ref{fig:ScoreTimeTrainGame} also shows that gaming experience correlates with average scores on these electrostatics assessments, including a pretest that was given before receiving any electrostatics instruction. Fig. \ref{fig:ScoreTimeTrainGame} shows that gamers scored around 13\% higher on all question types, except the final post in-VR questions, making this a much larger effect than any due to the VR treatment. This bolsters the possibility that the Fig.~\ref{fig:ScoreTrainGame} data showing untrained gamer students performing the worst in terms of gains, might be related to ceiling effects.

The original intent of this study was to compare overall gains between the two treatment groups, and not to compare groups' performances on individual questions or questions in one medium compared to another. It is therefore entirely possible that the observed uptick in performance on the 'post in-VR' questions is simply due to variation in question difficulty. It seems unlikely, though, that males could score 75\% on the pretest, and 77\% on the post in-VR questions if their difficulty levels were very different. The apparent equalization of male and female scores on the post in-VR questions thus strongly warrants additional study. 


Although we find evidence that VR training is beneficial for student acclimation to VR-based instruction, overall, we conclude that the VR-based instruction in this study has essentially no effect on student understanding overall. This is especially apparent if the final goal is for students to answer inherently 3D questions on 2D media like paper exams and computer screens. In this sense, our results agree with other large studies like Smith et al.\cite{smith17} (on electrostatics), Madden et al.\cite{madd18,madd19} (on moon phases), Porter et al.\cite{PERC19} (on magnetostatics), and Brown et al.\cite{brown19} (engineering) that VR-based instruction is not more effective than other media at teaching inherently 3D topics. Since interest in VR-based instruction in physics is unlikely to subside, one takeaway from our study (borne out particularly in Fig.~\ref{fig:ScoreTimeTrainGame}) is that VR training does seem to positively affect all students' ability to learn in a VR environment. This is true even for students who are not gamers, who, in our data, are more likely to be women.





The custom-designed assessment for electrostatics is adequately reliable for the purposes of this preliminary study, it has not been independently validated, and it could be improved in future work. A validated assessment would be especially useful for interpreting puzzling results like the differences we found between VR-based assessments of electrostatics knowledge versus assessments that were done on a typical computer. Having a validated instrument for electrostatics would clearly be a great benefit to the VR education community in much the same way that the Purdue Spatial Rotations test\cite{guay77} has been for other studies (e.g. Brown et al. \cite{brown19}).

\begin{acknowledgments}

The development of VR visualizations has been supported by the OSU STEAM Factory, OSU's Marion campus, and the Office for Distance Education and eLearning (ODEE).

\end{acknowledgments}

\bibliographystyle{unsrt}
\bibliography{main}

\end{document}